\documentclass[two column, prA]{revtex4}%
\usepackage{amsmath}
\usepackage{graphicx}
\usepackage{amsfonts}
\usepackage{amssymb}%
\providecommand{\U}[1]{\protect\rule{.1in}{.1in}}

\begin{document}
\title{Photon position measure}
\author{Margaret Hawton}
\email{margaret.hawton@lakeheadu.ca}
\affiliation{Department of Physics, Lakehead University, Thunder Bay, ON, Canada, P7B 5E1}

\begin{abstract}
The positive operator valued measure (POVM) for a photon counting array
detector is derived and found to equal photon flux density integrated over
pixel area and measurement time. Since photon flux density equals number
density multiplied by the speed of light, this justifies theoretically the
observation that a photon counting array provides a coarse grained measurement
of photon position. The POVM obtained here can be written as a set of
projectors onto a basis of localized states, consistent with the description
of photon position in a recent quantum imaging proposal [M. Tsang, Phys. Rev.
Lett. \textbf{102}, 253601 (2009)]. The wave function that describes a photon
counting experiment is the projection of the photon state vector onto this
localized basis. Collapse is to the electromagnetic vacuum and not to a
localized state, thus violating the text book rules of quantum mechanics but
compatible with the theory of generalized observables and the
nonlocalizability of an incoming photon.

\end{abstract}
\maketitle

\section{Introduction}

Although position of a massive particle is an observable in nonrelativistic
quantum mechanics, photon position is a controversial concept. It has been
argued that there is no photon number density, only energy density
\cite{Sipe}. Solutions to the photon wave equation are electric and magnetic
fields, but the relationship between these fields and the normalizable
Laudau-Peierls (LP) photon number amplitude is nonlocal \cite{LP,BB96}.
Localization of a converging (or diverging) photon pulse cannot be exact
since, according to the Paley-Weiner theorem, it must have subexponential
tails \cite{BB98}.

The most appropriate real space description of the photon is a subject of
ongoing debate. Since photons are detected through interaction with matter and
atomic dipoles are sensitive to electric field, the positive frequency part of
this field is often called the photon wave function \cite{Scully}. A LP-like
wave function has been defined in the context of the inverse problem,
spontaneous emission of a photon by an excited atom \cite{Eberly}. A
measurement of photon position should be described by projection onto a basis
of localized photon states \cite{HawtonLorentz}, but this probability
amplitude is not a positive frequency electromagnetic field.

In spite of this theoretical controversy, a single photon is counted in only
one pixel of an array detector and there is considerable interest in the
measurement of photon position in quantum optics. Position entanglement has
been demonstrated experimentally to be a way to achieve high dimensional
entangled states for information processing \cite{Boyd}. Recently Tsang
defined a photon POVM and applied it to a new quantum imaging method
\cite{Tsang}. The use of a POVM to describe photon position was introduced
first by Kraus \cite{Kraus}.

Use of a POVM bypasses the theoretical difficulties because a POVM is just an
operator partition of the identity that does not require a self-adjoint
position operator. Here a photon position POVM is obtained by lifting it to a
larger Hilbert space that includes the atoms of a photodetector. We will start
by considering absorption of photons by a two dimensional (2D) photon counting
array detector. This experiment is described by a projection valued measure
(PVM) that follows the usual rules for an observable in quantum mechanics.
Projection operators onto each pixel will be defined such that their
expectation values equal the probability to create an electron-hole pair. We
will then examine how this photon counting experiment gives information about
photon position.

\section{Theory}

In the theory of generalized observables a POVM is an operator partition of
the identity that describes the possible outcomes of a measurement. According
to Neumark's dilation theorem a POVM on the Hilbert space $\mathcal{P}$\ can
always be lifted to a PVM on a bigger Hilbert space, $\mathcal{H=P\otimes A}$,
where $\mathcal{A}$ is an ancillary subspace. The operators associated to a
measurement on $\mathcal{H}$, $\widehat{F}_{n}$, are then a set of
self-adjoint projection operators satisfying the usual postulates of quantum
mechanics. The POVM on $\mathcal{P}$ is the partial trace on the ancillary
Hilbert subspace, $\mathcal{A}$, that is%
\begin{equation}
\widehat{P}_{n}=Tr_{\mathcal{A}}\left(  \widehat{\rho}_{\mathcal{A}}%
\widehat{F}_{n}\right)  . \label{POVM}%
\end{equation}
This formalism will be applied to a photon counting measurement: In Subsection
A the localized states needed to define photon density will be discussed. In
B, time dependent photodetection operators will be derived, and in C the
special case of a thick detector that counts photons will be considered.
Limitation of the model are discussed in D, and in E and F the counting
operators will be applied to one and two photon states and collapse. Normal
incidence and the paraxial limit will be assumed so that cross talk between
pixels can be neglected.

\subsection{ Localized states}

A state with definite position is an equally weighted sum over all momenta.
States of this form were used by Tsang to describe photon position
\cite{Tsang}. Here periodic boundary conditions over the volume $V$ will be
used so that the momenta, $\mathbf{k}$, are discrete. Since all momenta are
included, the two transverse polarizations should be defined for all wave
vectors. In $\mathbf{k}$-space spherical polar coordinates the definite
helicity unit vectors $\mathbf{e}_{\mathbf{k},\sigma}^{(0)}=\left(
\widehat{\theta}+i\sigma\widehat{\phi}\right)  /\sqrt{2}$ can be defined for
$\sigma=\pm1$. In the paraxial approximation the basis states are then
circularly polarized (CP). The most general transverse basis vectors,
$\mathbf{e}_{\mathbf{k},\sigma}^{(\chi)}=\exp\left(  -i\sigma\chi\right)
\mathbf{e}_{\mathbf{k},\sigma}^{(0)}\left(  \theta,\phi\right)  ,$ are rotated
about $\mathbf{k}$ through the Euler angle $\chi\left(  \theta,\phi\right)  $
\cite{HawtonAM}. For example, if $\chi=-\phi$ the unit vectors are rotated
back to the $x$ and $y$-axes and describe linearly polarized (LP) photons in
the paraxial limit.

If the operator $\widehat{a}_{\mathbf{k},\sigma}$ annihilates a photon with
wave vector $\mathbf{k}$ and helicity $\sigma$, the operator
\begin{equation}
\widehat{a}_{\sigma}\left(  \mathbf{r},t\right)  =\frac{1}{\sqrt{V}}%
\sum_{\mathbf{k}}\widehat{a}_{\mathbf{k},\sigma}\exp\left(  i\mathbf{k\cdot
r}-ikct\right)  \label{Annihilation}%
\end{equation}
annihilates a photon with helicity $\sigma$ at position $\mathbf{r}$ and time
$t$. The one-photon localized states are%
\begin{equation}
\left\vert \mathbf{r},t,\sigma\right\rangle =\widehat{a}_{\sigma}^{\dagger
}\left(  \mathbf{r},t\right)  \left\vert \emptyset\right\rangle
\label{Localized}%
\end{equation}
where $\widehat{a}_{\sigma}^{\dagger}$ is the adjoint of $\widehat{a}_{\sigma
}$ and $\left\vert \emptyset\right\rangle $ is the electromagnetic vacuum
state. This is an orthonormal one-photon basis at any fixed time, $t$. The
number density operator is%
\begin{equation}
\widehat{n}\left(  \mathbf{r},t\right)  =\sum_{\sigma}\widehat{a}_{\sigma
}^{\dagger}\left(  \mathbf{r},t\right)  \widehat{a}_{\sigma}\left(
\mathbf{r},t\right)  . \label{Number}%
\end{equation}

\subsection{Photodetection}

It is assumed for concreteness that the photodetector is made up of a 2D array
of semiconductor pixels with the $z$-axis chosen parallel to its inward
normal. The dipole moment of the individuals atoms is taken to be isotropic
with components all equal to $\mu$. The atoms are assumed to lie on a lattice
so the atom positions are also discrete. The area, $A,$ and the time, $T,$
occupied by the electromagnetic pulse to be detected are taken to be finite so
that $V=cTA$. To obtain time dependent operators that describe photon
annihilation the Schr\"{o}dinger picture (SP) operators will be transformed to
the interaction picture (IP) and then to the Heisenberg picture (HP).

Glauber defined "an ideal photon detector as a system of negligible size
(e.g., of atomic or subatomic dimensions) which has a \ frequency-independent
photoabsorption probability" \cite{Glauber}. Detection of a photon with
polarization in the $p$-direction by a Glauber photodetector at $\mathbf{r}$
will be described by the SP projection operator $\left\vert e_{\mathbf{r}%
,p}\right\rangle \left\langle e_{\mathbf{r},p}\right\vert $ in the localized
Wannier basis satisfying $\left\langle e_{\mathbf{r},p}|e_{\mathbf{r}^{\prime
},p^{\prime}}\right\rangle =\delta_{\mathbf{r},\mathbf{r}^{\prime}}%
\delta_{p,p^{\prime}}$ \cite{KochHaug}. The factor $\delta_{p,p^{\prime}}$
describes absorption of a photon with definite polarization, as would be the
case for a spherically symmetric hydrogen-like atom making a transition from a
$ns$ to a $\left(  n+1\right)  p_{x},$ $\left(  n+1\right)  p_{y}$ or $\left(
n+1\right)  p_{z}$ state. For normal incidence in the paraxial approximation
we can choose $\widehat{\mathbf{z}}$ parallel to the direction of propagation
of the beam and $p=\sigma$ for any wave vector that is actually present in the
pulse. In this case we will only need $\widehat{a}_{\sigma}\left(
\mathbf{r},t\right)  $ defined by (\ref{Annihilation}), although more
generally $\widehat{a}_{\mathbf{k},p}=\sum_{\sigma}\widehat{a}_{\mathbf{k}%
,\sigma}\left(  \mathbf{e}_{\mathbf{k},\sigma}\cdot\mathbf{e}_{p}^{\ast
}\right)  $ where $\mathbf{e}_{p}$ is a mutually orthogonal triad of unit vectors.

The SP counting operators are%
\begin{equation}
\widehat{F}_{0}^{SP}=\widehat{1}_{\mathcal{A}}\otimes\widehat{1}_{\mathcal{P}%
}=\widehat{1} \label{0Photon}%
\end{equation}
for zero photons,%
\begin{equation}
\widehat{F}_{1,n}^{SP}=\sum_{p,\mathbf{r}\in D_{n}}\left\vert e_{\mathbf{r}%
,p}\right\rangle \left\langle e_{\mathbf{r},p}\right\vert \otimes\widehat
{1}_{\mathcal{P}} \label{1Photon}%
\end{equation}
for one photon in pixel $n$,
\begin{equation}
\widehat{F}_{2,n,n^{\prime}}^{SP}=\sum_{p,\mathbf{r}\in D_{n};p^{\prime
},\mathbf{r}^{\prime}\in D_{n^{\prime}}}\left\vert e_{\mathbf{r}^{\prime
},p^{\prime}}\right\rangle \left\vert e_{\mathbf{r},p}\right\rangle
\left\langle e_{\mathbf{r},p}\right\vert \left\langle e_{\mathbf{r}^{\prime
},p^{\prime}}\right\vert \otimes\widehat{1}_{\mathcal{P}} \label{2Photon}%
\end{equation}
for two photons, one in pixel $n$ and one in pixel $n^{\prime}\neq n$, and so
on. The sums are over polarization and all atoms in the $n^{th}$ pixel of the
photodetector and $\left\{  \mathbf{r}\in D_{n}\right\}  $ are the positions
of the atoms at which an elecron-hole (e-h) pair can be created. The unit
operator, $\widehat{1}_{\mathcal{P}}$, acting in the photon Hilbert subspace
$\mathcal{P},$ acknowledges the fact that no information is obtained directly
from the photons.

These operators will be transformed to the IP and then to the HP. The SP
Hamiltonian is
\begin{equation}
\widehat{H}^{SP}=\widehat{H}_{0}+\widehat{H}_{I}^{SP}+\widehat{H}%
_{I}^{SP\dagger} \label{H}%
\end{equation}
with zero order term $\widehat{H}_{0}$ and photon-atom interaction described
by%
\begin{equation}
\widehat{H}_{I}^{SP}=\mu\sum_{p,\mathbf{r}\in V}\widehat{E}_{\sigma}%
^{(+)SP}\left(  \mathbf{r}\right)  \left\vert e_{\mathbf{r},p}\right\rangle
\left\langle g\right\vert . \label{HIsp}%
\end{equation}
In (\ref{HIsp}), $\left\vert e_{\mathbf{r},p}\right\rangle \left\langle
g\right\vert $ creates an e-h pair at $\mathbf{r}$ with atomic dipole moment
in the $p$-direction. The SP positive frequency electric field operator is%
\begin{equation}
\widehat{E}_{p}^{(+)SP}\left(  \mathbf{r}\right)  =i\sum_{\mathbf{k}}E_{k}%
\exp\left(  i\mathbf{k\cdot r}\right)  \widehat{a}_{\mathbf{k},p} \label{Eop}%
\end{equation}
where, with $k\equiv\left\vert \mathbf{k}\right\vert $,%
\begin{equation}
E_{k}=\sqrt{\frac{\hbar kc}{2\epsilon_{0}V}}. \label{Ek}%
\end{equation}

In the IP and within the rotating wave approximation $\widehat{H}_{0}$ is
unchanged but the interaction Hamiltonian becomes time dependent. In a
semiconductor the atoms interact, so the Wannier states, $\left\vert
e_{\mathbf{r},p}\right\rangle $, do not have definite energy. The eigenvectors
of the atomic part of $\widehat{H}_{0}$ are the Bloch basis states,
$\left\vert \psi_{\mathbf{q},p}\right\rangle $, and its eigenvalues form a
band of energies, $\epsilon_{q}$, where $\epsilon_{q}$ equals to the gap
energy plus the energy of the electron hole pair. The $N$-atom Wannier basis
is related to the Bloch basis according to \cite{KochHaug}
\begin{equation}
\left\vert e_{\mathbf{r},p}\right\rangle =\frac{1}{\sqrt{N}}\sum_{\mathbf{q}%
}\exp\left(  i\mathbf{q}\cdot\mathbf{r}\right)  \left\vert \psi_{\mathbf{q}%
,p}\right\rangle . \label{e_r}%
\end{equation}
For a cubic lattice using periodic boundary conditions the wave vector
$\mathbf{q}$ has components $q_{i}=2\pi n_{i}/aN_{i}$ for $n_{i}=1$ to $N_{i}$
where $a$ is the interatomic distance and $N_{x}N_{y}N_{z}=N$ is the number of
atoms in the pixel. For the photon, the expansion (\ref{Eop}) gives states
with definite energy, $kc$. Thus in (\ref{HIsp}) $\widehat{a}_{\mathbf{k}%
,p}\left\vert \psi_{\mathbf{q},p}\right\rangle \left\langle g\right\vert $
acquires a time dependence $\exp\left[  i\left(  \epsilon_{q}-kc\right)
t\right]  $ in the IP \cite{ScullyCh6}, equivalent to the unitary
transformation $\widehat{O}^{IP}=\widehat{U}_{0}^{\dagger}\widehat{O}%
^{SP}\widehat{U}_{0}$ with
\begin{equation}
\widehat{U}_{0}\left(  t\right)  =\exp\left(  -i\widehat{H}_{0}t/\hbar\right)
. \label{U0}%
\end{equation}
We can transform from the IP to the HP by solving $i\hbar d\widehat
{U}/dt=\widehat{H}_{I}\widehat{U}$ iteratively to give the Dyson series
\cite{SakuraiModernQM}%
\begin{equation}
\widehat{U}\left(  t\right)  =\widehat{1}+\widehat{U}^{(1)}\left(  t\right)
+\widehat{U}^{(2)}\left(  t\right)  +... \label{U}%
\end{equation}
where%
\begin{equation}
\widehat{U}^{(1)}\left(  t\right)  =-\frac{i}{\hbar}\int_{t_{0}}^{t}%
dt^{\prime}\widehat{H}_{I}^{IP}\left(  t^{\prime}\right)  , \label{U1}%
\end{equation}%
\begin{equation}
\widehat{U}^{(2)}\left(  t\right)  =-\frac{1}{\hbar^{2}}\int_{t_{0}}%
^{t}dt^{\prime}\widehat{H}_{I}^{IP}\left(  t^{\prime}\right)  \int_{t_{0}%
}^{t^{\prime}}dt^{\prime\prime}\widehat{H}_{I}^{IP}\left(  t^{\prime\prime
}\right)  , \label{U2}%
\end{equation}
and it is assumed that the detector is in its ground state at time $t_{0}$.
The Hermitian conjugate has been omitted since photon emission by the
semiconductor will be neglected.

All photons incident on an ideal photon counting detector must be absorbed, by
definition of such a detector. Thus an $J$-photon counting operator must
reduce a $J$-photon state to the vacuum state, $\left\vert \emptyset
\right\rangle $. This is the case for terms of the form%
\begin{equation}
\widehat{F}_{J}^{HP}=\widehat{U}^{(J)\dagger}\left(  t\right)  \widehat{F}%
_{J}^{IP}\widehat{U}^{(J)\dagger}\left(  t\right)  . \label{FopHP}%
\end{equation}
In the HP the zero photon operator remains $\widehat{1}$, and the one-photon
operator is%
\begin{align}
\widehat{F}_{1,n}^{HP}\left(  t_{0}\right)   &  =\widehat{U}^{(1)\dagger
}\left(  t\right)  \widehat{F}_{1,n}^{IP}\left(  t\right)  \widehat{U}%
^{(1)}\left(  t\right) \nonumber\\
&  =\frac{\left\vert \mu\right\vert ^{2}}{\hbar^{2}}\sum_{p;\mathbf{r}%
,\mathbf{r}^{\prime},\mathbf{r}^{\prime\prime}\in D_{n}}\int_{t_{0}}%
^{t_{0}+\Delta t}dt^{\prime}\int_{t_{0}}^{t_{0}+\Delta t}dt^{\prime\prime
}\nonumber\\
&  \times\widehat{U}_{0}^{\dagger}\left(  t^{\prime}\right)  \left\vert
g\right\rangle \left\langle e_{\mathbf{r}^{\prime},p}\right\vert \widehat
{E}_{p}^{\left(  -\right)  SP}\left(  \mathbf{r}^{\prime}\right)  \widehat
{U}_{0}\left(  t^{\prime}\right)  \left\vert e_{\mathbf{r},p}\right\rangle
\left\langle e_{\mathbf{r},p}\right\vert \nonumber\\
&  \times\widehat{U}_{0}^{\dagger}\left(  t^{\prime\prime}\right)  \widehat
{E}_{p}^{\left(  +\right)  SP}\left(  \mathbf{r}^{\prime\prime}\right)
\left\vert e_{\mathbf{r}^{\prime\prime},p}\right\rangle \left\langle
g\right\vert \widehat{U}_{0}\left(  t^{\prime\prime}\right)  \label{F1}%
\end{align}
where $\Delta t$ is the time required for the measurement. The effect of the
unitary time development operators can be evaluated by transforming the
localized basis vectors, $\left\vert e_{\mathbf{r},p}\right\rangle $, to the
Bloch basis vectors that are delocalized within a pixel. The Wannier and Bloch
bases are both complete so $\sum_{\mathbf{r}}\left\vert e_{\mathbf{r}%
,p}\right\rangle \left\langle e_{\mathbf{r},p}\right\vert $ $\ $can be
replaced with $\sum_{\mathbf{q}}\left\vert \psi_{\mathbf{q},p}\right\rangle
\left\langle \psi_{\mathbf{q},p}\right\vert $ and (\ref{e_r}) can be
substituted in $\left\vert e_{\mathbf{r}^{\prime\prime},p}\right\rangle $ and
$\left\langle e_{\mathbf{r}^{\prime},p}\right\vert $. Using assumptions
equivalent to those in Kimble and Mandel's photodetection theory
\cite{KimbleMandel} in the impulsive detector sensitivity limit
\cite{Bondurant} and the orthogonality of the Bloch states, a factor
$\left\vert \int_{0}^{\infty}d\epsilon_{q}\exp\left(  -i\epsilon_{q}%
\tau\right)  \right\vert =2\pi\hbar\delta\left(  \tau\right)  $ is obtained.
With $\tau=t^{\prime\prime}-t^{\prime}$, Eq. (\ref{F1}) can then be written
as
\begin{equation}
\widehat{F}_{1,n}^{HP}\left(  t_{0}\right)  =s\int_{t_{0}}^{t_{0}+\Delta
t}dt\sum_{p,\mathbf{r}\in D_{n}}\widehat{E}_{p}^{\left(  -\right)  }\left(
\mathbf{r},t\right)  \widehat{E}_{p}^{\left(  +\right)  }\left(
\mathbf{r},t\right)  \left\vert g\right\rangle \left\langle g\right\vert
\label{F1n}%
\end{equation}
where $\widehat{E}_{p}^{\left(  +\right)  }\left(  \mathbf{r},t\right)  $ is
the IP positive frequency electric field operator, $s=2\pi\rho_{\epsilon
}\left\vert \mu\right\vert ^{2}/\hbar$ is the one-atom sensitivity to the
absolute square of the field per unit time, and $\rho_{\epsilon}$ is the
number of states per unit energy interval in the semiconductor. A Glauber
ideal photodetector has a flat frequency response, equivalent to assuming that
$s$ is constant, and we have followed that convention here. Absorption of
photons in different pixels is only correlated through the wave function, so
the multi-photon counting operators will be time ordered integrals of factors
of the form (\ref{F1n}).

Inside the detector the $z$-component of the wave vector acquires an imaginary
part, $i\alpha_{k}$, due to absorption, and the speed of light becomes $c/n$
where $n$ is the index of refraction. Frequency and the component of the wave
vector tangential to the interface is \ unchanged by entry into the detector.
If a mode with free space wave vector $\mathbf{k}$ approaches the detector at
an angle $\theta$ to its normal it is refracted to $\theta_{d}$ inside the
detector$\ $so that $\sin\theta_{d}=\sin\theta/n$ and $k_{z}\rightarrow
nk\cos\theta_{d}+i\alpha_{k}$. For normal incidence in the paraxial
approximation $\cos\theta_{d}=1.$ The IP positive frequency electric field
operator inside the photodetector is then%
\begin{align}
\widehat{E}_{p}^{\left(  +\right)  }\left(  \mathbf{r},t\right)   &
=i\sum_{\mathbf{k}}E_{k}\widehat{a}_{\mathbf{k},p}\nonumber\\
&  \times\exp\left(  ik_{x}x+ik_{y}y+inkz-ikct-\alpha_{k}z\right)  .
\label{Einside}%
\end{align}
\ 

\subsection{Photon counting}

The response of an ideal photon counting detector to photon flux is discussed
by Bondurant \cite{Bondurant}. The detector must be thick enough to absorb all
incident photons, so we can integrate over all $z$ inside the detector. To
integrate (\ref{F1n}) over $z$ we can convert the sum over atom positions to
an integral using $\sum_{\mathbf{r}}\rightarrow\rho_{a}\int d^{3}r$ where
$\rho_{a}$ is the density of atoms to give%
\begin{equation}
\widehat{F}_{1,n}^{HP}\left(  t\right)  =\int_{t_{0}}^{t_{0}+\Delta t}%
dt\int_{A_{n}}dxdy\widehat{w}_{1}\left(  x,y,t\right)  \left\vert
g\right\rangle \left\langle g\right\vert \label{Fn2counting}%
\end{equation}
where $A_{n}$ is the area of the $n^{th}$ pixel and%
\begin{equation}
\widehat{w}_{1}\left(  x,y,t\right)  =s\rho_{a}\sum_{p}\int_{0}^{\infty
}dz\widehat{E}_{p}^{\left(  -\right)  }\left(  \mathbf{r},t\right)
\widehat{E}_{p}^{\left(  +\right)  }\left(  \mathbf{r},t\right)
\label{PhotonOperator}%
\end{equation}
is the first-order coincidence rate operator in \cite{Glauber} and
\cite{Bondurant}.

The essence of Bondurant's calculation can be understood by consideration of a
single mode field with wave vector $\mathbf{k}$ and definite polarization.
Using (\ref{Einside}) and (\ref{Ek}), $\widehat{E}^{\left(  -\right)
}\widehat{E}^{\left(  +\right)  }=\hbar kc\widehat{a}^{\dagger}\widehat
{a}/2\epsilon_{0}V=\left(  \hbar kc/2\epsilon_{0}\right)  \widehat{n}$ where
$\widehat{n}=a^{\dagger}a/V$ decays inside the dielectric. Since $d\widehat
{w}_{1}\left(  \mathbf{r},t\right)  =\left(  s\rho_{a}\hbar kc/2\epsilon
_{0}\right)  $ $\widehat{n}dz$ is the absorption probability per unit area per
unit time over the distance $dz$, $d\widehat{n}=-d\widehat{w}_{1}/c=-\left(
s\rho_{a}\hbar k/2\epsilon_{0}\right)  $ $\widehat{n}dz$\ implies that
\begin{equation}
\alpha_{k}=\frac{s\rho_{a}\hbar k}{4\epsilon_{0}}. \label{alpha}%
\end{equation}
Using Equations (\ref{Ek}) and (\ref{Einside}), integration of
(\ref{PhotonOperator}) over a thick detector then gives $\widehat{w}_{1}%
=s\rho_{a}\int_{0}^{\infty}dz\widehat{E}^{\dagger}\widehat{E}=c\widehat{n}$,
that is energy density is replaced with number density, implying that a thick
photodetector counts photons.

More generally integration of (\ref{PhotonOperator}) over $z$ with
substitution of (\ref{Einside}) and (\ref{Ek}) gives \cite{Bondurant}%
\begin{align}
\widehat{w}_{1}\left(  x,y,t\right)   &  =\frac{s\rho_{a}\hbar c}%
{2\epsilon_{0}V}\sum_{p,\mathbf{k},\mathbf{k}^{\prime}}\frac{\sqrt{kk^{\prime
}}\widehat{a}_{\mathbf{k},p}^{\dagger}\widehat{a}_{\mathbf{k}^{\prime},p}%
}{\alpha_{k}+\alpha_{k^{\prime}}+in\left(  k-k^{\prime}\right)  }%
\label{PBoudurant}\\
&  \times\exp\left[  i\left(  k_{x}-k_{x}^{\prime}\right)  x+i\left(
k_{y}-k_{y}^{\prime}\right)  y\right. \nonumber\\
&  \left.  -i\left(  k-k^{\prime}\right)  ct\right]  .\nonumber
\end{align}
A photon should be detected with certainty if we sum over all pixels and count
for a time $T$. To test this we can evaluate $\sum_{n}\widehat{F}_{1,n}^{HP}$
for $t=0$ and $\Delta t=T$ . Substituting$\ $(\ref{PBoudurant}) in (\ref{F1n})
and summing over pixels we get a factor $\delta_{k_{x},k_{x}^{\prime}}%
\delta_{k_{y},k_{y}^{\prime}}\delta_{k,k^{\prime}}$ inside the $\mathbf{k}$
and $\mathbf{k}^{\prime}$ sum with the result
\begin{equation}
\sum_{n}\widehat{F}_{1,n}^{HP}=\sum_{p,\mathbf{k}}\widehat{a}_{\mathbf{k}%
,p}^{\dagger}\widehat{a}_{\mathbf{k},p}\left\vert g\right\rangle \left\langle
g\right\vert . \label{AllPhotons}%
\end{equation}
Thus a photon is counted with certainty if the detector is big enough and it
is operated over a long enough time.

The $k$-dependent factor in (\ref{PBoudurant}) can be expanded in powers of
the relative differential photon frequency%
\begin{equation}
\delta_{k}=\left(  k-k_{0}\right)  /k_{0} \label{delta}%
\end{equation}
where $k_{0}c$ is the center frequency of the pulse. To first order in
$\delta_{k}$%
\begin{align}
\widehat{w}_{1}\left(  x,y,t\right)   &  =\frac{c}{V}\sum_{p,\mathbf{k}%
,\mathbf{k}^{\prime}}\widehat{a}_{\mathbf{k},p}^{\dagger}\widehat
{a}_{\mathbf{k}^{\prime},p}\label{Poperator}\\
&  \times\exp\left[  i\left(  k_{x}-k_{x}^{\prime}\right)  x+i\left(
k_{y}-k_{y}^{\prime}\right)  y\right. \nonumber\\
&  \left.  -i\left(  k-k^{\prime}\right)  c\left(  t-\tau\right)  \right]
\nonumber
\end{align}
where $\tau$ is a small delay of the order of a few optical cycles
\cite{Bondurant}. This can be written as%
\begin{equation}
\widehat{w}_{1}\left(  x,y,t\right)  =c\widehat{n}\left(  x,y,0,t-\tau\right)
\label{PhotonCorrelation}%
\end{equation}
where $\widehat{n}\left(  \mathbf{r},t\right)  $ given by (\ref{Number}) is
the number density operator. This implies that a photon counting array
measures photon flux density. The factor $\exp\left(  -ik_{0}ct\right)  $
disappears from (\ref{PhotonCorrelation}), so $\widehat{n}$ describes the
envelope function of the pulse and is unaffected by its center frequency. The
$z$-integration is performed independently in each pixel and $t>t^{\prime}$
and $t<t^{\prime}$are equally likely, so the two-photon correlation operator
is%
\begin{equation}
\widehat{w}_{2}\left(  x,y,t;x^{\prime},y^{\prime},t^{\prime}\right)
=\frac{1}{2!}c^{2}\widehat{n}\left(  x^{\prime},y^{\prime},0,t^{\prime}%
-\tau\right)  \widehat{n}\left(  x,y,0,t-\tau\right)  .
\label{TwoPhotonCorrelation}%
\end{equation}

According to (\ref{POVM}), the one-photon POVM is the partial trace of
(\ref{Fn2counting}) over the atoms, $\widehat{P}_{1,n}=Tr_{\mathcal{A}}\left(
\widehat{\rho}_{\mathcal{A}}\widehat{F}_{1,n}^{HP}\right)  $. For a
photodetector initially in its ground state $\widehat{\rho}=\widehat{\rho
}_{\mathcal{P}}\otimes\left\vert g\right\rangle \left\langle g\right\vert $ so
that the trace operation just eliminates $\left\vert g\right\rangle
\left\langle g\right\vert $ giving%
\begin{equation}
\widehat{P}_{1,n}\left(  t_{0}\right)  =\int_{t_{0}}^{t_{0}+\Delta t}%
dt\int_{A_{n}}dxdy\widehat{w}_{1}\left(  x,y,t\right)  \text{.}
\label{PhotonPOVM}%
\end{equation}
If two photons are absorbed the POVM is%
\begin{align}
\widehat{P}_{2,n,n^{\prime}}\left(  t_{0}\right)   &  =\int_{t_{0}}%
^{t_{0}+\Delta t}dt\int_{A_{n}}dxdy\label{TwoPhotonPOVM}\\
&  \times\int_{t_{0}}^{t}dt^{\prime}\int_{A_{n^{\prime}}}dx^{\prime}%
dy^{\prime}\widehat{w}_{2}\left(  x,y,t;x^{\prime},y^{\prime},t^{\prime
}\right)  .\nonumber
\end{align}
These are coarse-grained photon position operators. The pixel location can be
identified with the $x$ and $y$ position of the photon, while the $z$ position
of the photon is determined by the measurement time.

\subsection{Limitations of the model}

The photon counting array considered here will not count photons in all beams
configurations. In the extreme case of propagation away from the detector, no
photons will be counted at all. For modes not normal to the detector surface
some photons will be counted in pixels adjacent to their entry pixel, and
spatial resolution will be reduced. To go beyond the paraxial approximation or
for application to beams not normal to the detector surface, (\ref{Einside})
and equations derived from it would have to be generalized.

A real photodetector has a frequency dependent quantum efficiency less that
one that should be corrected for when the photon number is determined from the
number of electron-hole pairs created. If it is not possible to determine
whether more than one photon was absorbed in a particular pixel, the photon
density should be low enough that multiple photon absorption in a single pixel
is unlikely.

It was assumed in (\ref{1Photon}) that an absorbed photon has definite
polarization. This was proved to be the case for normal incidence with $E_{z}$
neglected in \cite{KimbleMandel}. In general, the Hermitian tensor $Tr\left\{
\widehat{\rho}\widehat{E}_{p}^{(-)}\left(  \mathbf{r},t\right)  \widehat
{E}_{p^{\prime}}^{(+)}\left(  \mathbf{r},t\right)  \right\}  $ can be
diagonalized so "there always exist a set of three (complex) orthogonal
polarization vectors such that the field components in these directions are
statistically uncorrelated" \cite{Glauber}. If we use these polarization
vectors as a basis, the diagonal form is exact.

To first order in an expansion over the relative differential photon frequency
a thick photodetector counts photons with a small delay time of a few optical
cycles. The delay $\tau$ has a simple physical interpretation. It is due to
the $z$-integral that samples the square of the field over a distance of a few
penetration depths, and these photons arrived at an earlier time.

Second order terms in the $\delta_{k}$ expansion that go as $\delta_{k}^{2}$
and $\delta_{k}\delta_{k^{\prime}}$ result in $\omega_{0}^{-2}\widehat
{a}\left(  \partial^{2}\widehat{a}/\partial t^{2}\right)  $ and $\omega
_{0}^{-2}\left(  \partial\widehat{a}/\partial t\right)  \left(  \partial
\widehat{a}/\partial t\right)  $ in real space where $\omega_{0}=ck_{0}.$
There is also a term that goes as $\delta_{k}\left(  k_{x}^{2}+k_{y}%
^{2}\right)  /k_{0}^{2}$ that gives $\omega_{0}^{-1}k_{0}^{-2}\left(
\partial\widehat{a}/\partial t\right)  \left(  \partial^{2}\widehat
{a}/\partial x^{2}+\partial^{2}\widehat{a}/\partial y^{2}\right)  $ in real
space. These higher order terms depend on the rate of change of the pulse
envelope relative to the optical frequency and are negligible in most experiments.

For a single atom the factor $\sqrt{k}$ gives a nonlocal relationship between
field and number amplitude such that the field at $\mathbf{r}^{\prime}$ of a
photon at $\mathbf{r}$ is proportional to $\left\vert \mathbf{r}%
-\mathbf{r}^{\prime}\right\vert ^{-7/2}$ \cite{Amrein,BB96}. This nonlocality
is almost completely eliminated by integration over $z$, and is neglected here.

\subsection{Photon state and wave function}

As an illustration (\ref{PhotonCorrelation}) and (\ref{TwoPhotonCorrelation})
will be applied to paraxial one and two photon states at normal incidence. For
a one-photon pulse, if the time required for a measurement is less than the
pulse length, detection of either $1$ or $0$ photons is possible. The
probability to count one photon between times $t_{0}$ and $t_{0}+\Delta t$ is
$P_{1}=\sum_{n}P_{1,n}$ given by (\ref{PhotonPOVM}) while the zero-photon
probability is $1-P_{1}$.\ If this intermediate photon number is read out, the
wave function collapses to a $0$ or $1$ photon state \cite{Ueda}. To avoid the
complications of intermediate collapse we will only consider measurement times
greater than the pulse length.

A HP one-photon pure state is of the form%
\begin{equation}
\left\vert \psi_{1}\right\rangle =\sum_{\mathbf{k},\sigma}c_{\mathbf{k}%
,\sigma}\left\vert \mathbf{k},\sigma\right\rangle \label{OnePhoton}%
\end{equation}
where $\sum_{\mathbf{k},\sigma}\left\vert c_{\mathbf{k},\sigma}\right\vert
^{2}=1.$ The expectation value of the photon flux density operator is%
\begin{align}
&  \left\langle \psi_{1}\left\vert \widehat{w}_{1}\left(  x,y,t\right)
\right\vert \psi_{1}\right\rangle \nonumber\\
&  =c\sum_{\sigma}\left\langle \psi_{1}\left\vert \widehat{a}_{\sigma
}^{\dagger}\left(  x,y,t\right)  \right\vert 0\right\rangle \left\langle
0\left\vert \widehat{a}_{\sigma}\left(  x,y,t\right)  \right\vert \psi
_{1}\right\rangle .
\end{align}
The photon wave function will be defined here as the projection of $\left\vert
\psi_{1}\right\rangle $ onto the one-photon state (\ref{Localized}) with
polarization $\sigma$ localized at $\mathbf{r}$ at the measurement time $t$,
that is%
\begin{align}
\psi_{\sigma}\left(  \mathbf{r},t\right)   &  =\left\langle \mathbf{r}%
,t,\sigma|\psi_{1}\right\rangle \nonumber\\
&  =\sum_{\mathbf{k}}c_{\mathbf{k},\sigma}\frac{\exp\left(  -i\mathbf{k}%
\cdot\mathbf{r}+ikct\right)  }{\sqrt{V}}. \label{WaveFunction}%
\end{align}
At each point in space $\psi_{\sigma}$ has two components corresponding to the
two transverse photon polarizations. For a CP paraxial beam the unit vectors
at $\theta=0$ reduce to $\mathbf{e}_{\pm1}^{(0)}=\left(  \widehat{\mathbf{x}%
}+i\sigma\widehat{\mathbf{y}}\right)  /\sqrt{2}$ and, together with
$\widehat{\mathbf{z}}$, can be identified with the $\mathbf{e}_{p}$ basis. For
LP we can take $\chi=-\phi$ so that $\mathbf{e}_{1}^{(-\phi)}=\widehat
{\mathbf{x}}$ and $\mathbf{e}_{-1}^{(-\phi)}=\widehat{\mathbf{y}}$ \ at
$\theta=0$. In either case
\begin{equation}
\left\langle \psi_{1}\left\vert \widehat{w}_{1}\left(  x,y,t\right)
\right\vert \psi_{1}\right\rangle =c\sum_{\sigma}\left\vert \psi_{\sigma
}\left(  x,y,0,t\right)  \right\vert ^{2}, \label{w1Sigma}%
\end{equation}
is photon flux density.

A two-photon HP state vector is of the form%
\begin{equation}
\left\vert \psi_{2}\right\rangle =\sum_{\mathbf{k},\mathbf{k}^{\prime}%
,\sigma,\sigma^{\prime}}c_{\mathbf{k},\sigma;\mathbf{k}^{\prime}%
,\sigma^{\prime}}\left\vert \mathbf{k},\sigma;\mathbf{k}^{\prime}%
,\sigma^{\prime}\right\rangle . \label{TwoPhoton}%
\end{equation}
If we define the symmetric two-photon wave function%
\begin{align}
\psi_{\sigma,\sigma^{\prime}}\left(  \mathbf{r},t;\mathbf{r}^{\prime
},t^{\prime}\right)   &  =\frac{1}{\sqrt{2!}}\left\langle \emptyset\right\vert
\widehat{a}_{\sigma}\left(  \mathbf{r},t\right)  \widehat{a}_{\sigma^{\prime}%
}\left(  \mathbf{r}^{\prime},t^{\prime}\right)  \left\vert \psi_{2}%
\right\rangle \label{2PhotonPsi}\\
&  =\frac{1}{\sqrt{2!}V}\sum_{\mathbf{k},\mathbf{k}^{\prime}}\left(
c_{\mathbf{k},\sigma;\mathbf{k}^{\prime},\sigma^{\prime}}+c_{\mathbf{k}%
^{\prime},\sigma^{\prime};\mathbf{k},\sigma}\right) \nonumber\\
&  \times\exp\left[  i\left(  -\mathbf{k}\cdot\mathbf{r-k}^{\prime}%
\cdot\mathbf{r}^{\prime}+kct+k^{\prime}ct^{\prime}\right)  \right]  ,\nonumber
\end{align}
the two-photon coincidence rate is
\begin{align}
&  \left\langle \psi_{2}\left\vert \widehat{w}_{2}\left(  x,y,t;x^{\prime
},y^{\prime},t^{\prime}\right)  \right\vert \psi_{2}\right\rangle \nonumber\\
&  =c^{2}\sum_{\sigma,\sigma^{\prime}}\left\vert \psi_{\sigma,\sigma^{\prime}%
}\left(  x,y,0,t;x^{\prime},y^{\prime},0,t^{\prime}\right)  \right\vert ^{2}.
\label{w2sigma}%
\end{align}
For example, if the photons are produced by spontaneous parametric down
conversion (SPDC) it was found that their positions are highly correlated
\cite{Boyd}. For type II SPDC the polarization of the photons is orthogonal,
while for type I the polarizations are the same. Both these effects can be
described by (\ref{w2sigma}). In the photon counting experiment considered
here polarization was not measured, and should be summed over. Knowledge of
photon polarization comes from knowledge of the beam properties rather than
from the photodetection process.

\subsection{Collapse}

We will start by considering a pure state, $\left\vert \psi_{1}\right\rangle
\left\vert g\right\rangle $, that initially contains one-photon and assume
that an e-h pair has been detected in pixel $n$. The extended system of
photons plus detector atoms satisfies the usual rules of quantum mechanics. In
the IP this state vector evolves with time to $\left\vert \Psi\left(
t\right)  \right\rangle =\widehat{U}^{(1)}\left(  t\right)  \left\vert
\psi_{1}\right\rangle \left\vert g\right\rangle $ and collapses to the final
state $\left\vert \Psi_{1,f}^{IP}\right\rangle =\widehat{F}_{1,n}%
^{IP}\left\vert \Psi\left(  t\right)  \right\rangle $ after a measurement.
Since its final and initial states are the same, $\widehat{F}_{1,n}%
^{IP}=\widehat{F}_{1,n}^{SP}$ is given by (\ref{1Photon}) so that
\begin{equation}
\left\vert \Psi_{1,f}^{IP}\right\rangle \propto\sum_{\sigma=\pm1,\mathbf{r}\in
D_{n}}\left\vert e_{\mathbf{r},\sigma}\right\rangle \left\langle
e_{\mathbf{r},\sigma}\right\vert \widehat{U}^{(1)}\left(  t\right)  \left\vert
\psi_{1}\right\rangle \left\vert g\right\rangle .
\end{equation}
If $\psi_{\sigma}\left(  x,y,0,t\right)  $ is uniform over the area of a pixel
and polarization isn't measured the normalized collapsed state vector
simplifies to%
\begin{equation}
\left\vert \Psi_{1,f}^{IP}\right\rangle =\frac{1}{\sqrt{2N}}\sum
_{\sigma,\mathbf{r}\in D_{n}}\left\vert e_{\mathbf{r},\sigma}\right\rangle
\left\vert \emptyset\right\rangle
\end{equation}
where $N$ is the number of atoms in the $n^{th}$ pixel. If $J$ photons are
detected and the wave function is approximately uniform over a pixel the
system collapses to the state
\begin{equation}
\left\vert \Psi_{J,f}^{IP}\right\rangle =%
{\displaystyle\prod\limits_{i=1}^{J}}
\left(  \frac{1}{\sqrt{2N}}\sum_{\sigma_{i},\mathbf{r}_{i}\in D_{n_{i}}%
}\left\vert e_{\mathbf{r}_{i},\sigma_{i}}\right\rangle \right)  \left\vert
\emptyset\right\rangle .
\end{equation}

\section{Discussion}

The one-photon flux density operator (\ref{PhotonCorrelation}) for helicity
$\sigma$ can be written as $c\left\vert \mathbf{r},t,\sigma\right\rangle
\left\langle \mathbf{r},t,\sigma\right\vert $ evaluated at $z=0$. The
Hermitian position operator $\widehat{\mathbf{r}}_{\sigma}=\int d^{3}%
r\mathbf{r}\left\vert \mathbf{r},t,\sigma\right\rangle \left\langle
\mathbf{r},t,\sigma\right\vert $ with eigenvectors $\left\vert \mathbf{r}%
,t,\sigma\right\rangle $ and eigenvalues $\mathbf{r}$ and $\sigma$ can be
defined such that the usual rules of quantum mechanics are formally satisfied.
For two photons the projection operators can be written as $\left\vert
\mathbf{r},t,\sigma;\mathbf{r}^{\prime},t^{\prime},\sigma^{\prime
}\right\rangle \left\langle \mathbf{r},t,\sigma;\mathbf{r}^{\prime},t^{\prime
},\sigma^{\prime}\right\vert $, consistent with the 1D multi-photon POVMs
defined by Tsang \cite{Tsang}. There are in principle many ways to construct a
POVM, since any partition of the identity is allowed. However, an exactly
localized position eigenvector is an equally weighted sum over all wave
vectors, and this form arose naturally here. But there is an important
deviation from the usual rules; collapse is not to this position eigenvector.

By the usual rules of quantum mechanics, if position of a photon with
polarization $\sigma$ is measured at time $t$ the system should collapse to
the instantaneously localized state, $\left\vert \mathbf{r},t,\sigma
\right\rangle $. Since all wave vectors are equally likely in a localized
state such as (\ref{Localized}), both incoming and outgoing states are
required \cite{HawtonLorentz}. In a photon counting experiment there is no
outgoing wave and collapse to an exactly localized state is not possible. In
place of the usual rules, here we have used the theory of generalized
observables that separates the calculation of probabilities from collapse.
With this formalism we found that a photon counting array detector measures
coarse grained photon position, while collapse is to the zero photon vacuum
state. Some theorists would insist we say "absorption position" was measured,
but the term "photon position" is widely used.

Definite helicity basis states were selected because all wave vectors can be
included and helicity is Lorentz invariant. States with a definite
polarization direction in real space exclude longitudinal wave vectors and
thus can not be localized. The definite helicity unit vectors have definite
total angular momentum parallel to some $z$-axis \cite{HawtonAM}. Selecting
the $z$-axis is analogous to selecting the quantization axis for electron spin
that is arbitrary in the absence of symmetry breaking, but can be fixed by a
magnetic field. For the photon, symmetry can be broken by the beam axis.

In nonrelativistic quantum mechanics the real space wave function is both
$\left\langle \mathbf{r}|\psi\right\rangle $ and a solution to the particle's
wave equation. For photons these are not the same thing, the former being a
scalar while the latter is a vector whose $\mathbf{k}^{th}$ term is weighted
as $\sqrt{k}$. Here the photon wave function is defined in the former sense,
that is as the projection of the photon state vector onto a basis of the
localized states. The photon wave function is $\psi_{\sigma}\left(
\mathbf{r},t\right)  =\left\langle \mathbf{r},t,\sigma|\psi_{1}\right\rangle $
and the photon flux density measured in a photon counting experiment is
$c\sum_{\sigma}\left\vert \psi_{\sigma}\left(  x,y,0,t\right)  \right\vert
^{2}$. The quantum electrodynamic state vector is used to describe the photon
state and a wave function that satisfies Maxwell's equations is not required.
This choice of wave function is consistent with Chan et al \cite{Eberly}, with
Tsang \cite{Tsang}, and with the discussion of position eigenvectors in
\cite{HawtonLorentz}.

In summary, a \ photon position POVM for a photon counting experiment has been
derived by performing a partial trace over the ancillary Hilbert space of the
photodetector atoms. It was found that the photon counting probability equals
the photon flux density integrated over the area of the pixel and the time
taken for a measurement. This is, of course, as expected, since a photon
counting array is designed to measure photon position. If the photon wave
function is defined as the projection of its state vector onto the localized
states, photon flux density is its absolute square multiplied by the speed of
light. The photons are destroyed in this measurement, but the distribution of
positions and times at which they are detected is a measure of their
probability density in the incoming pulse.

\textit{Acknowledgements: }The author thanks the Natural Sciences and
Engineering Research Council for financial support.

\end{document}